# REDUCING COMPETITIVE CACHE MISSES IN MODERN PROCESSOR ARCHITECTURES

Milcho Prisagjanec and Pece Mitrevski

Faculty of Information and Communication Technologies,
University "St. Kliment Ohridski", Bitola, Republic of Macedonia

## ABSTRACT

*The increasing number of threads inside the cores of a multicore processor, and competitive access to the shared cache memory, become the main reasons for an increased number of competitive cache misses and performance decline. Inevitably, the development of modern processor architectures leads to an increased number of cache misses. In this paper, we make an attempt to implement a technique for decreasing the number of competitive cache misses in the first level of cache memory. This technique enables competitive access to the entire cache memory when there is a hit – but, if there are cache misses, memory data (by using replacement techniques) is put in a virtual part given to threads, so that competitive cache misses are avoided. By using a simulator tool, the results show a decrease in the number of cache misses and performance increase for up to 15%. The conclusion that comes out of this research is that cache misses are a real challenge for future processor designers, in order to hide memory latency.*

## 1. INTRODUCTION

The main objective in the development of a processor is its performance increase, and the bottlenecks are being eliminated by implementation of different types of techniques in the architecture itself. By following this timeline historically, one can note that the frequency increase, the implementation of out-of-order execution of the instruction stream, the enlargement of the instruction window, and the Instruction-Level Parallelism (ILP), all contributed to increasing the performances of processors in some periods of time [1].

But, according to some authors, the gap between processor and memory speeds, which has existed since the very appearance of computers, despite many offered techniques (some of which have been already implemented in commercial processors), is the reason for performance decline [2]. An instruction that accesses memory, until it is complete with memory data, blocks the processor resources at the same time and thus decreases performance. When the processor needs memory data, and data are found in the first level cache, one says that there is a hit. Otherwise, a cache miss occurred and the memory system starts a procedure of elimination, which may take several hundred cycles [3].

The main reasons for performance decline and increased number of cache misses in modern processor architectures are the memory shared among cores, as well as the drift towards implementation of larger number of cores and threads in the processor itself. Some of the techniques that are intended to help in reducing cache misses (prefetching, for example), even contribute to the appearance of such competitive cache misses and decrease performance.

 49

The aim is to find a technique which would enable for reducing memory latency and the number of cache misses in commercial processors. There are many papers in which authors contribute to the increase of Memory-Level Parallelism, but none of these has been implemented in commercial processors to date. Others go even further, stating that it is necessary to abandon Instruction-Level Parallelism and that researchers should completely turn to Memory-Level Parallelism instead [4].

Taking that into consideration, after we review related work in Section 2 and analyze the performance impact of competitive cache misses in Section 3, we propose a technique for reducing competitive cache misses. The suggestion is every single thread to have its own "virtual" cache memory in the first level during block replacement, while using the entire cache memory during loading. On the positive side are the easy way of implementation, and the fact that it does not "revolutionary" change the present architecture of the processor. But, on the negative side, one should note that the cache memory is used far below par in the first level during not very intensive processor workload, as we will demonstrate in Section 4. Section 5 concludes the paper.

## 2. RELATED WORK

The research made by many authors showed that cache misses reduce the processor performance even up to 20%. Even more significant information is that the gap between the processor and memory speeds is getting deeper and bigger through time, which contributes to a larger number of cache misses. This was the main motivation for various efforts to design techniques that reduce cache misses. For example, the prefetching technique [5], which has been implemented in commercial processors, supplies memory data by fetching them into the cache memory before the processor needs them. Some other authors [6] have given ideas of a virtual enlargement of the instruction window, so that it would be possible to unlock the processor's resources in the window and to load in advance memory data which are to be used by the processor in the near future. They provide a technique for virtual extension of the instruction window, which is called runahead execution. The idea is to enable continued operation of the processor when its resources are blocked by a long latency instruction. In this manner, memory data prefetching from the first level cache is enabled.

The idea for the improvement of processor performance by implementing Memory-Level Parallelism (MLP) techniques has been used in a number of research papers. Models that reduce memory latency and/or the number of cache misses are given herewith. Continual Flow Pipelines [7] aim to remove the long latency instruction together with its dependent instructions, so that resources would be released and the processor could execute the independent instructions. In [8], the author suggests a Recovery-Free Value Prediction technique, whose idea relates to speculative execution in order to load memory data from the first level cache. Out-of-order commit [9] is a technique which enables the execution of the instructions in the absence of a reordering buffer. In other words, by increasing the instruction window, there is a need for a constant increase of the reordering buffer which, in turn, increases the complexity of the processor. This technique allows to completely leave out the reordering buffer.

## 3. COMPETITIVE CACHE MISSES AND PERFORMANCE IMPACT

The lack of ability to burst through the technological cutbacks in processor design, changes the course in their development. That means that the idea is to design an architecture which would perform multiple numbers of instructions at the same clock frequency. The first-hand concept is based on a system in supercomputers known beforehand, where many computers execute their workload in parallel. This architecture is modified by using more processor cores placed on one silicon crystal. Each of the cores is a processor itself, but the new architecture significantly reduces energy dissipation. The trends of increasing the number of cores and threads are more present in modern processor architectures – they impose the necessity of a larger instruction window. By doing so, there is an increasing number of instructions which have greater need of memory data, whose latency is too big and causes stagnation in the work of the processor.

### 3.1. PERFORMANCES OF MULTI CORE PROCESSORS

In the best case, dual core processor should complete a programmer's code twice as fast, compared to a single core processor. But the practical cases so far showed that it does not really happen, and that the double core processor is only 1.5 times faster than a single core processor. Surprisingly, in some scenarios, single core processors even show better performances – one of the reasons are cache misses.

In [10], by means of a simulator, a single core processor has been designed. First level (L1) of cache memory is composed of instructional cache and data cache (2 x 64 Kbytes), L2 cache memory with a size of 256 Kbytes and L3 cache memory with a size of 512 Kbytes. The remaining parameters are configured according to the parameters of the commercial processors. The tests have been performed in architectures of one-level, two-level, and three-level cache memory. The results of the experimental tests are shown in Table 1. They show that there are misses at every level of cache memory: in L1 cache memory there are 88% hits and 12% misses. From the misses in L1 cache memory, 8.7% are next level cache memory misses (i.e. L2), whereas 5.3% of these misses are from the last (L3) cache memory level. The analysis shows that with the development of the new commercial architectures of processors and with the increase in the number of cores, there is also an increase in the number of cache misses, which reduce the performances of the processor.

Table 1.  Cache misses in presence of memory hierarchy.

| Architecture of cache memory | Cache memory | | | |
|---|---|---|---|---|
| | Number of accesses | Cache misses in L1 | Cache misses in L2 | Cache misses in L3 |
| 2x64+256+512 | 20000 | 2560 | 224 | 12 |

An analysis of the cache misses has been done depending on the number of the cores. By using a simulator, a multicore processor has been designed with only L1 cache memory with capacity of 512 Kbytes (i.e. 256 Kbytes of instruction cache and 256 Kbytes of data cache). The bus supports MESI protocol for maintaining memory coherence. The main memory capacity is 1 GB. Benchmarks of the designed processor have been executed by changing the number of processor cores. The results are shown in Fig. 1.

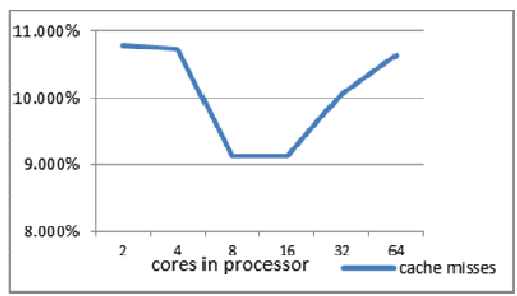

Figure 1. How the number of processor cores affects cache misses

One can see that the number of cache misses for the simulated architecture is 10% of the number of memory accesses. But it can also be concluded that by increasing the number of cores above 16, the number of cache misses increases as well, which has a negative effect on processor performance. One of the reasons for that is the competitive cache misses.

## 3.2. COMPETITIVE CACHE MISSES

In modern processor architectures, a number of cores and threads share the same memory, and the workload that the processor has to undertake is divided into processes and threads. Every process needs certain memory data, and if it cannot provide them from the first level cache, it loads them from the memory subsystem by using a certain method for changing the blocks. This methods do not usually check if a block, which is currently being changed, would be used in near future by some other processes. Namely, it leads to removal of some memory blocks, which might be needed by some other active processes in subsequent cycles. As a result, it would cause a decrease in processor performance.

This situation would become even worse if prefetching was used to load memory data from the first level cache. If every process would use this technique to load memory data from the first level of cache memory, which it would need in near future, due to the restricted capacity of the cache memory and its shared use, would cause competitive memory access and mutual removal of the blocks of data for the processes. The possibility of appearance of the competitive cache memory access grows with processor workload, with the increasing of the number of cores and threads that share the same memory, and finally because of the restricted capacity of the cache memory of the processor.

## 4. REDUCING COMPETITIVE CACHE MISSES – TECHNIQUE AND BENEFITS

To avoid competitive access of the threads, we suggest the technique of a one-way shared cache memory. The first level cache is shared to as many parts as the number of threads. All threads competitively access the memory. If a thread contains an instruction that accesses the memory, a check is performed whether the required memory data are available. Therefore, if these data can be found in the first level cache, then they are being loaded. So far, it is the same as in all the modern processor architectures. If the data cannot be found in the first level of the cache memory, then it is necessary to be loaded from another level of the memory subsystem. The loading of the memory block is completed by using one of the known replacement techniques, but by doing so the block can be loaded only in the part of the cache memory, which belongs to the thread. That means that, during loading memory data, the thread uses all the shared cache memory from the first level. But if certain data do not exist in the first level, then it loads them from the memory subsystem, but only in its own "virtual" part of the cache memory.

The prefetching technique, which is frequently used in modern processor architectures, is another reason for the appearance of competitive cache misses. Principally, this technique is implemented in order to decrease memory latency and the number of cache misses. In order to get a picture for the benefits of using the data prefetching technique, a processor has been tested and the number of cache misses depending on the number of prefetched data has been determined by using a simulator tool. In Fig. 2 it can be seen how the number of cache misses changes according to the number of prefetched instructions – in the interval of 4 to 6 prefetched instructions the number of cache misses is very close to 0.

With the results in this part we have further confirmed the efficiency of data prefetching in reducing the number of cache misses. However, this efficiency is optimal only if the prefetching system makes a correct prediction of the memory data that should be fetched and when to start the fetching process. This means that, for the instructions that access the memory, this technique makes early loading of memory data and places them in the first level cache. At the moment of execution of the instruction the memory data are already in the cache memory and the processor just takes them from there with minimum delay. But, in modern processor architectures, there is competitive access of the threads to the first level cache with the prefetching technique. That is the reason for competitive access to the same memory address while loading the memory data from the upper levels of the cache memory.

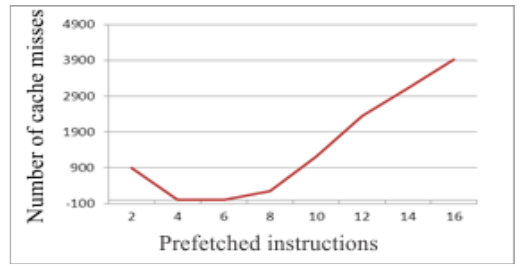

Figure 2. Number of cache misses depending on the number of prefetched instructions

## 4.1. SIMULATION ENVIRONMENT

In order to get measurable and comparable performance evaluation results, we designed a processor simulator by using the Python programming language [11]. SimPy (Simulation in Python) [12] is an object-oriented, process-based discrete-event simulation language based on standard Python, which provides the modeler with components of a simulation model including "Processes" (active components) and "Resources" (passive components) and provides monitor variables to assist in gathering statistics. Discrete-event simulation (DES) utilizes a mathematical/logical model of a physical system that portrays state changes at precise points in simulation time. Both the nature of the state changes and the time at which the changes occur, require precise description. Within DES, time advances not at equal size time steps, but rather until the next event can occur, so that the duration of activities determines how much the clock advances. "Process Execution Methods" (PEMs) use the "yield hold" command to temporarily delay process objects' operations. Figs. 3-7 show the program codes of the "RAM Memory" and "Instruction Pointer" resources, as well as the "Decoding", "Execution" and "Cache" processes, respectively.

```
ram=Place() # RAM Memory
for i in range(0,len(ip.buffer_ins)):
    if ip.buffer_ins[i][1]==True:
        entry=ip.buffer_ins[i][0],ip.buffer_ins[i][1],ip.buffer_ins[i][2]
        ram.buffer_ins.append(entry)
```

Figure 3. Program code of the "RAM Memory" resource in SimPy

```
ip=Place() #Instruction pointer
for i in range(1000,Num_ins,1):
    if random.random() < 0.225:  #22.5% instructions to access memory
        access=True
        address=int(random.triangular(1,500))
    else:
        access=False
        address=0
    entry=i,access,address
    ip.buffer_ins.append(entry)
```

Figure 4. Program code of the "Instruction Pointer" resource in SimPy

```
class Decoding(Process):
    def __init__(self):
        Process.__init__(self)
        self.time_exe=0.4  # time for execution
        self.pipeline=4    # change pipelined in processor
        self.capacity=32   # capacity of instruction windows

    def izvrsi(self):
        while True:
            number_ins=1
            for item in ip.buffer_ins:
                if len(cpu.buffer_ins) > self.capacity and self.capacity!=0:
                    break
                cpu.buffer_ins.append(ip.buffer_ins[0])
                del ip.buffer_ins[0]
                number_ins=number_ins+1
                if number_ins > self.pipeline:
                    break
                yield hold, self, abs(random.normalvariate(self.time_exe,1))
```

Figure 5. Program code of the "Decoding" process in SimPy

```
class Execution(Process):
    def __init__(self):
        Process.__init__(self)
        self.time_exe=0.4 # time for execution
        self.pipeline=8 # pipelined in processor
        self.max_capacity=0 #  maximum capacity of instruction window

    def izvrsi(self):
        while True:
            number_ins=0
            # execution of instructions without access to memory
            for cpu_inst in cpu.buffer_ins:
                if cpu_inst[1]==False:
                    rob.buffer_ins.append(cpu_inst)
                    cpu.buffer_ins.remove(cpu_inst)
                    number_ins=number_ins+1
            # execution of instructions with access to memory. If the data is in cache
            for cpu_inst in cpu.buffer_ins:
                if True in [cpu_inst[2] in x for x in kes1.buffer_ins]:
                    rob.buffer_ins.append(cpu_inst)
                    cpu.buffer_ins.remove(cpu_inst)
                    number_ins=number_ins+1

            yield hold, self, abs(random.normalvariate(self.time_exe,0.5))
            if self.max_capacity < len(cpu.buffer_ins):
                self.max_capacity=len(cpu.buffer_ins)
            if number_ins > self.pipeline:
                break
            if len(rob.buffer_ins)==Num_ins-1000:
                stopSimulation()
                print  'number of cache misses' , cache_missis
                print  'maximum capacity' ,        self.max_capacity
                print  'end program' ,             ,now()
```

Figure 6. Program code of the "Execution" process in SimPy

```
class Cache(Process):
    def __init__(self):
        Process.__init__(self)
        self.time2_exe=2.5 # time for execution Cache2-Cache1
        self.time3_exe=10 # time for execution Cache3-Cache2
        self.time4_exe=60 # time for execution Ram-Cache3
        self.time_exe=1
        self.num_inst=1
        self.mem_pipeline=1

    def izvrsi(self):
        while True:
            for instruction in cpu.buffer_ins:
                if instruction[1]==True:
                    if not True in [instruction[2] in x for x in Cache1.buffer_ins] :# If miss in cashes 1
                        global cache_missis
                        cache_missis=cache_missis+1
                        self.num_inst=self.num_inst+1
                        if True in [instruction[2] in x for x in Cache2.buffer_ins]:# If hit in cache 2:
                            #Fetch data from cache2 to cache1
                            if len(Cache1.buffer_ins) < Cache1_cap:
                                Cache1.buffer_ins.append(instruction)
                            else:
                                del Cache1.buffer_ins[0]
                                Cache1.buffer_ins.append(instruction)
                            self.time_exe=self.time_exe+abs(random.normalvariate(self.time2_exe,0.5))
                        else: # If miss in cache 2
                            if True in [instruction[2] in x for x in Cache3.buffer_ins]:  # If hit in cashes 3
                                #Fetch data from cache3 to cache2 and cache2
                                if len(Cache2.buffer_ins) < Cache2_cap:
                                    Cache2.buffer_ins.append(instruction)
                                else:
                                    del Cache2.buffer_ins[0]
                                    Cache2.buffer_ins.append(instruction)
                                if len(Cache1.buffer_ins) < Cache1_cap:
                                    Cache1.buffer_ins.append(instruction)
                                else:
                                    del Cache1.buffer_ins[0]
                                    Cache1.buffer_ins.append(instruction)
                                self.time_exe=self.time_exe+abs(random.normalvariate(self.time3_exe,0.5))
                            else: # If miss in cash 3
                                #Fetch data from RAM to cache3, cache2 and cache1
                                if len(Cache3.buffer_ins) < Cache3_cap:
                                    Cache3.buffer_ins.append(instruction)
                                else:
                                    del Cache3.buffer_ins[0]
                                    Cache3.buffer_ins.append(instruction)
                                if len(Cache2.buffer_ins) < Cache2_cap:
                                    Cache2.buffer_ins.append(instruction)
                                else:
                                    del Cache2.buffer_ins[0]
                                    Cache2.buffer_ins.append(instruction)
                                if len(Cache1.buffer_ins) < Cache1_cap:
                                    Cache1.buffer_ins.append(instruction)
                                else:
                                    del Cache1.buffer_ins[0]
                                    Cache1.buffer_ins.append(instruction)
                                self.time_exe=self.time_exe+abs(random.normalvariate(self.time4_exe,0.5))
                        if self.num_inst > self.mem_pipeline:
                            break
                #print self.time_exe
                yield hold, self, self.time_exe
                self.time_exe=1
#Start of the simulation
Num_ins=1000+20000
Cache1_cap=128
Cache2_cap=256
Cache3_cap=512
```

Figure 7. Program code of the "Cache" process in SimPy

## 4.2. PERFORMANCE BENEFITS

Next we tested the performances of an ideal processor without cache misses, a multithreaded processor, and a processor where we implemented the proposed technique for non-competitive access to the cache memory in the first level. The results gained from the tests are shown in Fig. 8. The offered technique is implemented in a single core multithreaded processor with 4 threads. The simulator makes parallel execution of the instructions in all of the threads. When the simulator comes to an instruction with access to the memory, it checks whether data can be found in the L1 cache memory. If there is a hit, it loads memory data and completes the instruction. But, if a miss in L1 cache memory occurs, the simulator searches these memory data in the upper levels of the memory hierarchy. The simulator uses the technique of block replacement and loads the memory data only in the virtual part of L1 cache memory, which belongs to the matching thread.

According to the results shown, the technique offered makes the time of execution of a program shorter for 15%. The source of this finding is the decreased number of competitive cache misses in the first level cache. The good side of the offered technique is that it does not "revolutionary" change the architecture of the processors. It can be implemented very easily: the change comes only in the technique of block replacement in the first level cache. A tag is used for marking which block belongs to which thread. While replacing the memory block, this tag is used to virtually divide cache memory by the number of threads that access the memory.

The prefetching technique is not something novel: the change only means that this technique early loads memory data in the second level of the cache memory, to avoid competitive access to the first level. It does not increase processor complexity, and so, it does not cause additional energy dissipation. The disadvantages of this technique, however, are the low exploitation of the first level cache during a poor workload of the processor. In cases with a small number of processes that reduces the number of active threads, as well, it effectively reduces the available capacity of the first level cache.

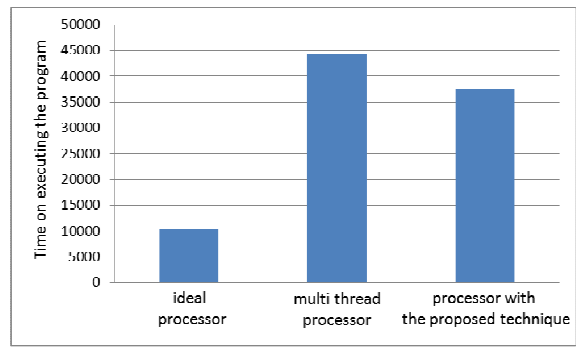

Figure 8. Graphical overview of the performances of three types of processors

## 5. CONCLUSIONS

The simulation results showed again the huge impact of cache misses on processor performance – the decreased processor performance due to cache misses cannot be ignored. The offered technique succeeds in decreasing the number of cache misses for about 15%, and it increases processor performance. Different studies have come to a conclusion that the number of cache misses in modern processor architectures is still outsized. On the other side, the direction in which they develop (by increasing the number of cores and threads, and by using shared cache memory), brings about the appearance of bigger number of cache misses even more, and decreases processor performance. Our research certified again that the cache misses significantly reduce performances, and that decreasing the number of cache misses is a great challenge to future research and processor designers.

**Authors**

**Milcho Prisagjanec** received his BSc degree in Electrical Engineering from the Faculty of Electrical Engineering and Information Technologies at the Ss. Cyril and Methodius University in Skopje, and the MSc degree from the Faculty of Technical Sciences, University "St. Kliment Ohridski" – Bitola, Republic of Macedonia. He is currently a system administrator and a PhD student at the Faculty of information and Communication Technologies. His field of interest is performance and reliability analysis of computer and communications systems.

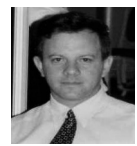

**Pece Mitrevski** received his BSc and MSc degrees in Electrical Engineering and Computer Science, and the PhD degree in Computer Science from the Ss. Cyril and Methodius University in Skopje, Republic of Macedonia. He is currently a full professor and Dean of the Faculty of Information and Communication Technologies, University "St. Kliment Ohridski" – Bitola, Republic of Macedonia. His research interests include Computer Architecture, Computer Networks, Performance and Reliability Analysis of Computer Systems, e-Commerce, e-Government and e-Learning. He has published more than 100 papers in journals and refereed conference proceedings and lectured extensively on these topics. He is a member of the IEEE Computer Society and the ACM.

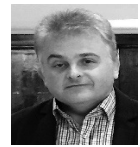